\documentclass{sigchi}
\usepackage{hyperref}
\usepackage{balance}       
\usepackage{graphics}      
\usepackage[T1]{fontenc}   
\usepackage{txfonts}
\usepackage{mathptmx}
\usepackage{color}
\usepackage[table]{xcolor}
\usepackage{booktabs}
\usepackage{textcomp}
\usepackage{enumitem}

\usepackage{microtype}        
\usepackage{ccicons}          

\usepackage{todonotes}
\usepackage{dblfloatfix}
\usepackage{fixltx2e}
\usepackage{paralist}

\def\plaintitle{SIGCHI Conference Proceedings Format}

\def\emptyauthor{}
\def\plainkeywords{Authors' choice; of terms; separated; by
  semicolons; include commas, within terms only; required.}

\makeatletter
\def\url@leostyle{%
  \@ifundefined{selectfont}{
    \def\UrlFont{\sf}
  }{
    \def\UrlFont{\small\bf\ttfamily}
  }}
\makeatother
\def\pprw{8.5in}
\def\pprh{11in}

\definecolor{linkColor}{RGB}{6,125,233}
\hypersetup{%
  pdftitle={\plaintitle},
  pdfauthor={\emptyauthor},
  pdfkeywords={\plainkeywords},
  pdfdisplaydoctitle=true, 
  bookmarksnumbered,
  pdfstartview={FitH},
  colorlinks,
  citecolor=black,
  filecolor=black,
  linkcolor=black,
  urlcolor=linkColor,
  breaklinks=true,
  hypertexnames=false
}

\begin{document}

\title{Boomerang: Rebounding the Consequences\\ of Reputation Feedback on Crowdsourcing Platforms}

 \numberofauthors{1}
\author{  
\alignauthor{ 
Snehalkumar (Neil) S. Gaikwad, Durim Morina, Adam Ginzberg, \\Catherine Mullings, Shirish~Goyal, Dilrukshi Gamage, Christopher Diemert, \\Mathias Burton, Sharon Zhou, Mark~Whiting, Karolina Ziulkoski, Alipta Ballav, \\Aaron Gilbee, Senadhipathige S. Niranga, Vibhor Sehgal, Jasmine Lin, Leonardy Kristianto, \\Angela Richmond-Fuller,  Jeff Regino,  Nalin~Chhibber, Dinesh Majeti, Sachin Sharma, \\Kamila Mananova, Dinesh Dhakal,  William Dai,  Victoria Purynova, Samarth Sandeep,\\ Varshine Chandrakanthan, Tejas Sarma, Sekandar Matin, Ahmed Nasser, Rohit Nistala, \\Alexander Stolzoff, Kristy Milland, Vinayak Mathur, Rajan Vaish, Michael S. Bernstein﻿ \\ \vspace{0.5em}
\affaddr{Stanford Crowd Research Collective  \\ Stanford University}\\
\email{daemo@cs.stanford.edu}}\\
}


\maketitle
\begin{abstract}

Paid crowdsourcing platforms suffer from low-quality work and unfair rejections, but paradoxically, most workers and requesters have high reputation scores. These inflated scores, which make high-quality work and workers difficult to find, stem from social pressure to avoid giving negative feedback. We introduce Boomerang, a reputation system for crowdsourcing that elicits more accurate feedback by rebounding the consequences of feedback directly back onto the person who gave it. With Boomerang, requesters find that their highly-rated workers gain earliest access to their future tasks, and workers find tasks from their highly-rated requesters at the top of their task feed. Field experiments verify that Boomerang causes both workers and requesters to provide feedback that is more closely aligned with their private opinions. Inspired by a game-theoretic notion of incentive-compatibility, Boomerang opens opportunities for interaction design to incentivize honest reporting over strategic dishonesty.

\end{abstract}
\category{H.5.3.}{Group and Organization Interfaces}{}
\begin{figure}[tb] 
\centering
  \includegraphics[width=0.7\columnwidth]{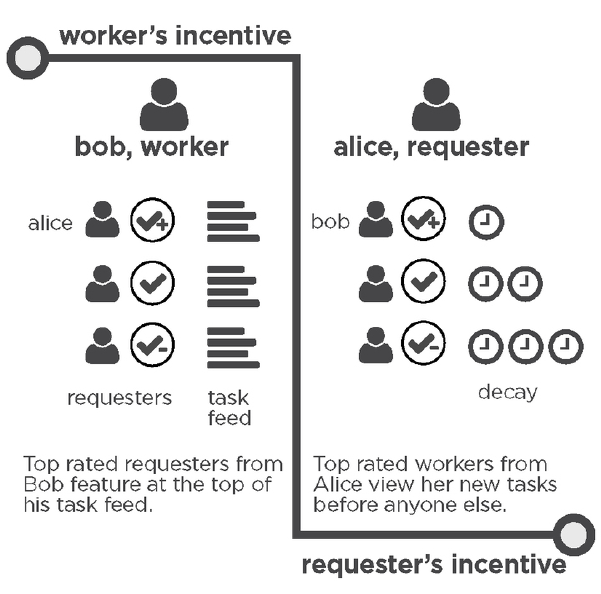}
  \caption{Boomerang is a reputation system that rebounds the consequences of ratings back onto the people who gave them. With Boomerang, requesters' ratings determine which workers get early access to their tasks, and workers' ratings determine the ranking of their task feed. Boomerang produces an incentive to rate accurately: falsely inflating a ranking will lead to a poorer experience.}
  \label{fig:boomerangincentive}
\end{figure}

\section{Introduction}
Today's crowdsourcing platforms are markets for lemons~\cite{akerlof1970market,ipeirotis2010mechanical}. On crowdsourcing platforms such as Amazon Mechanical Turk and Upwork, both workers and requesters face significant uncertainty: workers struggle to predict whether a requester will pay for their work~\cite{martin2014being, irani2013turkopticon}, and requesters worry whether workers will produce high-quality results~\cite{kittur2008crowdsourcing, mason2012conducting}. To reduce this uncertainty, crowdsourcing platforms rely on reputation systems such as task acceptance rates for workers and star ratings for requesters~\cite{irani2013turkopticon}. However, reputation systems are unable to address this uncertainty: for example, workers' acceptance rates on Mechanical Turk are often above 97\%, regardless of the quality of their work~\cite{mitra2015personorprocess}. The result is a downward spiral~\cite{akerlof1970market}: requesters offer lower wages to offset their risk of low-quality results, and workers respond to lower payment with lower quality work~\cite{ipeirotis2010mechanical, bederson2011web, silberman2010sellers}. Ultimately, ineffective reputation systems lead to dissatisfaction, lost productivity, and abandonment.

\begin{figure*}
\centering
  \includegraphics[width=2.0\columnwidth]{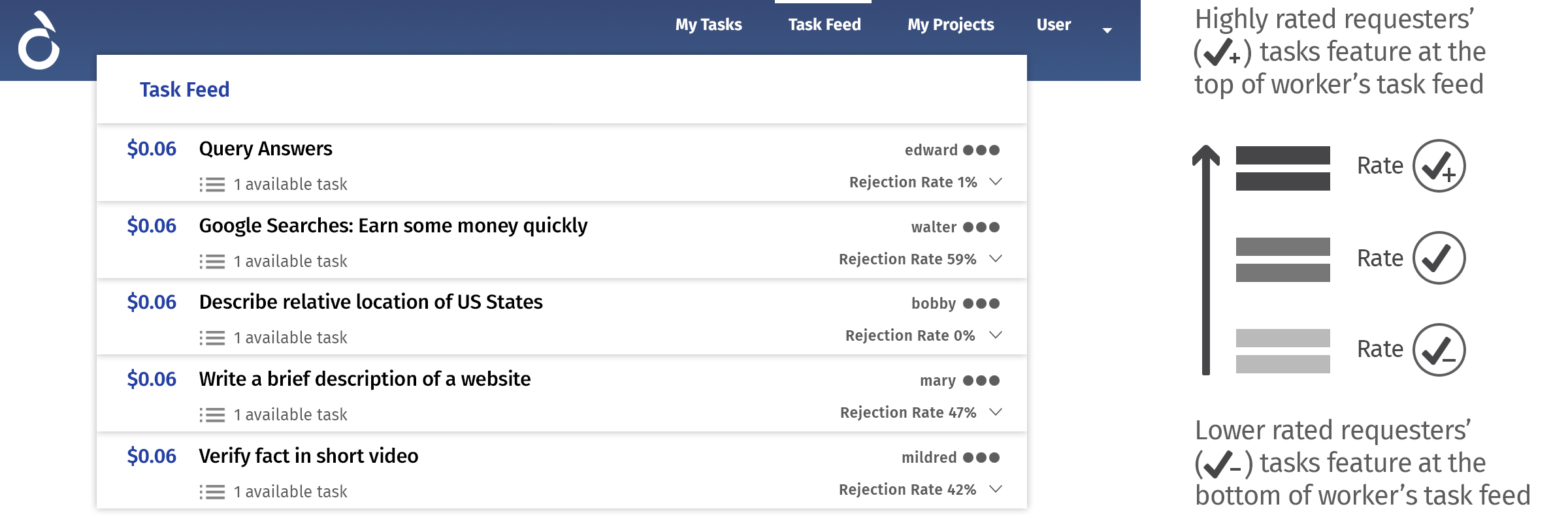}
  \caption{Daemo's task feed shows a list of tasks from which workers can select. Boomerang places requesters whom the worker has rated highly (\checkmark+) at the top, and those whom the worker has rated poorly (\checkmark-) at the bottom.
}~\label{fig:taskFeed}
\end{figure*}

Reliable reputation information does exist---workers and requesters form detailed opinions about each other as they interact---but this information is difficult to elicit accurately. The core challenge is \emph{reputation inflation}: significant social costs to providing negative feedback cause people to ``generously round up'' their feedback scores, even if they hope to never work with that partner again~\cite{horton2015reputation, ipeirotis2010quality, silberman2010sellers}. Incentives are misaligned: for example, by the time a worker realizes that a requester is low-quality, the worker is already mid-way through the task and has little reason to report the issue accurately. For their part, many requesters never reject work---even terrible work---because they do not want to deal with the resulting complaints~\cite{navarro2015ethical}. Algorithms can derive reputation information post-hoc by comparing workers' responses to each other~\cite{sheng2008get} and to ground truth~\cite{aroyo2013goldstandard}, but tasks that are open-ended cannot utilize these techniques. As a result, workers and requesters often regress to manual curation of trusted partners~\cite{mitra2015personorprocess, turkAlert}. If accurate reputation information is desired~\cite{kittur2013future}, no current design elicits it. 

In this paper, we draw on \textit{incentive compatibility}~\cite{nisan2007algorithmic, kamar2012incentives} as an interaction design strategy to capture more accurate reputation feedback for crowdsourcing systems. Incentive compatibility, in economic game theory, requires a set of rules in which people maximize their own outcomes by sharing their private information truthfully. As a result, incentive compatibility has traditionally been applied to economic environments where it is possible to influence people's behaviors under a formal set of rules and expected monetary payoffs~\cite{jurca2007collusion} such as crowdsourcing payment schemes~\cite{singla2013truthful,singer2013pricing}. In this paper, we explore whether the core ideas behind incentive compatibility can apply more broadly to the wicked~\cite{rittel1973dilemmas} socio-technical design of a crowdsourcing system. We pursue incentive compatible interaction design by introducing designs in which people benefit the most from the system when they truthfully rate and share honest opinions of their partners. In contrast to traditional reputation systems, which allow people to leave feedback with no impact on themselves, an incentive-compatible design might find a way to rebound the consequences of inaccurate feedback back onto the person who gave it.

We draw on incentive-compatible interaction design strategies to introduce \textit{Boomerang}, a reputation system for crowdsourcing marketplaces. In order to align workers' and requesters' incentives when leaving reputation feedback, Boomerang takes users' feedback and then ``boomerangs'' it back so that their feedback impacts their activities later in the platform, such that inaccurate reporting results in a lower-quality experience and accurate reporting results in a higher-quality experience for requesters and workers: 

\begin{compactenum}
\item Rating: When a requester posts a new task, Boomerang grants early access to workers whom the requester rated highly in the past (Figure~\ref{fig:boomerangincentive}). Thus, inflating the rating of a low-quality worker increases the probability that the low-quality worker will complete that requester's future tasks before others can, reducing the quality of future work results. Similarly, Boomerang ranks workers' task feeds using their ratings, making it easier to find requesters who they have rated highly (Figure~\ref{fig:taskFeed}).

\vspace{1em}
\item Rejection: Boomerang displays a personalized estimate of each requester's rejection rate to workers in the task feed. For example, a high-rated worker sees a requester's rejection rate for other high-rated workers, while a low-rated worker sees that requester's rejection rate for other low-rated workers. Aware that workers will avoid requesters with high rejection rates, requesters become incentivized to avoid attracting low-quality workers by never rejecting work, or repelling high-quality workers by frequently rejecting work.
\end{compactenum}

We evaluate Boomerang through a series of field experiments. The first set of  experiments provide evidence that Boomerang causes workers and requesters to provide rating feedback that is more closely aligned with their private opinions. The second set of experiments demonstrate that workers react to the visibility of rejection information, and knowing this, requesters become more accurate at accepting and rejecting submissions. 

So far, we have implemented Boomerang by integrating it with the Daemo crowdsourcing platform~\cite{gaikwad2015daemo}. However, Boomerang introduces the opportunity to explore incentive-compatible interaction design on crowdsourcing platforms, and on socio-technical platforms more broadly, by encouraging contributors to share privately-held information accurately and effectively. If successful, incentive-compatible interaction design could help these platforms build stronger foundations of trust and more effective contribution models.

\section{Related work}
Boomerang draws on research on crowdsourcing marketplaces as well as literature on mechanism design and incentive compatibility.

\subsection{Unreliable signals on crowdsourcing platforms}
Paid crowdsourcing platforms frame an ideal of connecting requesters with distributed human intelligence, while providing workers with opportunities to supplement their income, enhance their skills, and find longer-term career opportunities~\cite{kittur2013future}. In pursuit of this, platforms have attracted a large set of diverse workers (e.g.,~\cite{gupta2012mclerk, narula2011mobileworks}). Prior research has focused on recruiting and guiding crowd work through new channels such as local communities~\cite{heimerl2012communitysourcing}, experts~\cite{retelny2014expert}, short bursts of free time~\cite{Vaish:2014}, workflows inspired by distributed computing~\cite{kittur2011crowdforge}.

However, paid crowdsourcing platforms still struggle to ensure high-quality results for requesters and fair payment for workers, threatening the stability of these socio-technical ecosystems~\cite{kittur2013future, martin2014being, salehi2015we, brawley2016work}. While many issues are at fault, two in particular are a result of unreliable information shared on the platform: a) inflated reputation scores, which make it difficult to anticipate whether a worker or requester is truly 
high-quality~\cite{horton2015reputation,kittur2013future}, and b) unpredictable requester rejection behavior, which creates uncertainty around task acceptance~\cite{kittur2013future, martin2014being}. These challenges lead to downstream issues with requester activities such as pricing tasks~\cite{Justin2015ETA,ChrisCallison2014crowdworkers} and determining worker accuracy~\cite{sheng2008get}. 

This paper argues that one source of these challenges is the misaligned incentives between requesters and workers. For example, in Amazon Mechanical Turk's reputation system, workers are filtered based on their acceptance rates and prior task performance~\cite{mitra2015personorprocess}, but the requesters who assign these scores rarely have a future stake in those workers' reputation being accurate. Due to this misalignment, reputation information on crowdsourcing platforms is often an \textit{externality}, meaning that those who benefit or suffer are different from those who engage in the behavior~\cite{buchanan1962externality}. Boomerang aims to address these problems by causing users to internalize the downstream effects of their behaviors via the design of the reputation and rejection mechanisms of crowdsourcing platforms.

\subsection{Rating and rejection} 
Whether through acceptance rates (e.g., Amazon Mechanical Turk) or star ratings (e.g., Upwork), online reputation~\cite{resnick2002trust} is the main filter used by both workers~\cite{irani2013turkopticon} and requesters~\cite{christoforaki2015system, mason2012conducting} on crowdsourcing platforms. However, designing a crowdsourcing reputation system to elicit fair and accurate ratings remains a challenge. Public feedback is highly inflated: there is social pressure to rate highly so that workers and requesters can continue to be competitive in the marketplace, and to avoid any possibility of retaliation~\cite{Adamic11ratingfriends, horton2015reputation}. Private feedback can mitigate this problem~\cite{Teng2010Rate,horton2015reputation}, but does not erase it. This challenge motivates our design of a system that incentivizes workers and requesters to tell the truth about the quality of their partners.

Creating psychological nudges, payment structures, and game-theoretical incentives can help encourage high-quality crowd work~\cite{kamar2012incentives, shaw2011designing, law2011human, zhang2012reputation}. However, these incentives do not reciprocally apply to requesters' rejection behavior: requesters still hold immense power to deny payment for finished tasks~\cite{martin2014being}. To address this challenge, we designed Boomerang's rejection strategy to influence requesters to reject and accept tasks more fairly.

\subsection{Incentives and truth-telling}
Boomerang draws heavily on the concept of incentive compatibility as a motivation for its design. Incentive compatibility, a game-theoretical stance in which behaving truthfully is the best possible (or \textit{dominant}) strategy, plays a vital role in defining people's empirical truth-telling behavior~\cite{nisan2007algorithmic, papaioannou2005incentives}. Market-oriented computing research such as advertising auctions make liberal use of the concept, for example with Vickrey-Clarke-Groves (VCG) auctions~\cite{nisan2007algorithmic}, which are one of the most popular auctions in use today~\cite{edelman2005internet, FacebookAd}. Similarly, single item second-price auctions are won by the top-bidding agent, but that agent pays the amount of the second-highest bid. This mechanism incentivizes agents to bid honestly, rather than inflating or lowballing their bid. They achieve the best outcome for themselves by bidding honestly regardless of another's actions~\cite{nisan2007algorithmic}. Inspired by this work, we designed an incentive structure that motivates workers and requesters to share private information about the quality of their partners.

\section{Boomerang}   
In this section, we explore incentive-compatible interaction design through Boomerang (Figure~\ref{fig:boomerangincentive}), a reputation system for crowdsourcing platforms that incentivizes users to rate honestly and to provide their privately-held information accurately. We begin this section by introducing our goal of incentive-compatible interaction design. Following this introduction, we present Boomerang's two components: rating feedback and rejection transparency.

Because Boomerang requires significant infrastructural changes to existing closed-source crowdsourcing platforms, we have implemented it as part of the open-source Daemo platform~\cite{gaikwad2015daemo}. Daemo includes worker task feed and task completion interfaces, requester project authoring and result review interfaces, and other basic features of a paid crowdsourcing platform. On Daemo, tasks are the basic unit of work, and a series of tasks are grouped into a single unit called a project. Daemo cross-posts projects to Amazon Mechanical Turk when workers are unavailable.

\subsection{Incentive-compatibility in interaction design}
Crowdsourcing is no stranger to game-theoretic analyses (e.g.~\cite{pickard2011time,singer2013pricing}.) This prior research takes place in scenarios where agents are \textit{rational} and \textit{self-interested}---meaning they are trying to maximize only their own outcomes, such as earnings for a worker or result quality for a requester. To influence agents' decisions, the system designers create a \textit{mechanism}, or a set of rules and payoffs. Because agents are rational, they may act strategically or lie if the mechanism makes it advantageous for them to do so. For a mechanism to be \textit{incentive-compatible}, acting truthfully must lead to the best outcome for every agent~\cite{nisan2007algorithmic}. VCG auctions are one well-known incentive-compatible mechanism: designers create a system whereby agents have no incentive to deviate from a truth-telling strategy, hence preventing the system from collapsing under antisocial strategic behavior.

Boomerang seeks a design for \textit{wicked} socio-technical problems~\cite{rittel1973dilemmas}---in which agents do not share a clearly-definable strategy and the set of possible behaviors are unknown---such as those seen on crowdsourcing platforms. In crowdsourcing, even money is insufficient as a sole incentive: money is important, but it is not a total determinant of behavior~\cite{antin2012social,mason2009financial}. However, the designer has a wide variety of interaction designs that they might utilize. We thus generalize from the precise game-theoretic definition of incentive compatibility to pursue \textit{incentive-compatible interaction design}, in which the designer influences users to share honest information with the system. 

\textit{We seek socio-technical designs where the user benefits more by sharing honest information with the platform than by sharing inaccurate information}. The interaction design must result in a significantly less useful system if a user acts dishonestly or manipulatively. In this light, designs such as collaborative recommender systems~\cite{resnick1994grouplens} and the ESP Game~\cite{von2008designing, law2011human} aim to be incentive compatible, because strategic or dishonest behavior will only lead to worse recommendations or lower scores for the user~\cite{jain2008game}. Other designs, such as dating sites~\cite{hancock2007truth} and the UC San Diego DARPA Shredder Challenge team~\cite{stefanovitch2014error}, represent non-incentive compatible systems, because a contributor can share information dishonestly or strategically with no cost to themselves. Similar to crowdsourcing reputation systems, these systems must instead assume a norm of honest behavior. And like the Shredder Challenge, crowdsourcing marketplaces have witnessed this shared assumption broken and difficult to repair.

The following two designs represent our exploration into incentive-compatible interaction design for crowdsourcing. To begin, we rebound the consequences of inaccurate reputation ratings back onto the person who made them.

\subsection{Rating feedback}
In bilateral rating systems, such as those on crowdsourcing platforms where workers and requesters rate each other, there is significant social pressure to inflate ratings and ``generously round up'' in order to avoid retaliation, acquiesce to complaints, or enable someone continue to work on the platform~\cite{horton2015reputation}. In such situations, the strategic decision for workers and requesters is to rate their counterpart higher than they might privately believe. This strategic behavior inflates the overall ratings on the platform and leads to uncertainty about each person's true quality.

As workers submit work on Daemo, or after they receive feedback, they have the option to rate requesters on a 3-point scale.\footnote{Multidimensional or five-point rating systems would function similarly. We began with a traditional 5-point rating system~\cite{irani2013turkopticon}, but feedback from workers indicated that a 3-point model of bury/keep/love would be clearer.} These ratings correspond to \checkmark- (below expectations), \checkmark\ (meets expectations), and \checkmark+ (exceeds expectations). Likewise, when requesters review workers' submissions, they may rate the workers. This is a holistic rating system where requesters may, for example, be assigned one rating that accounts for task clarity, fair payment, and communication.
  
In Boomerang, ratings are designed to directly affect workers' and requesters' future work partners. The ratings that a requester gives to workers determine which workers will gain early access to the requester's future tasks. Likewise, the ratings that a worker gives to requesters determines the ranking of tasks in their task feed, with tasks from high-rated requesters at the top and those from low-rated requesters at the bottom. Each effect is intended to counter social pressure by introducing an individual incentive: requesters want higher quality work, and workers want to quickly find higher quality requesters~\cite{chilton2010task}. If the individual incentive is sufficiently strong, each user will truthfully share their private information about the quality of their counterparts. Below, we describe the design in detail.

\subsubsection{Requesters' incentive: workers' access to tasks}
\begin{figure}[tb]
\centering
  \includegraphics[width=1.0\columnwidth]{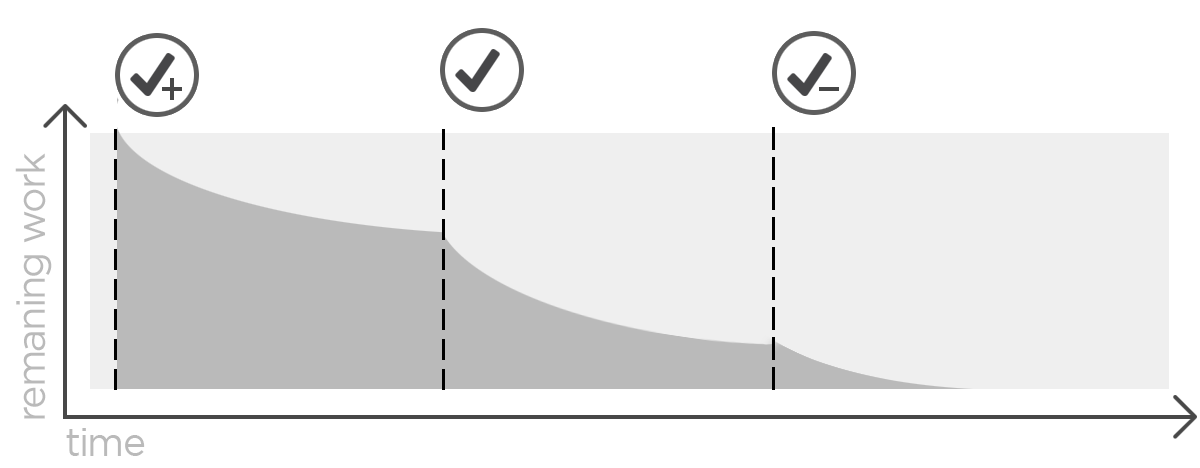}
  \caption{Boomerang cascades task release to \checkmark+ workers first. When their completion rate slows, it opens the task to \checkmark\ workers, then finally \checkmark- workers.}
  \label{fig:cascade}
\end{figure}

On current platforms, requesters typically have little incentive to rate workers accurately (if at all), in part because those ratings have no bearing on their future result quality. Boomerang uses the ratings that a requester gives a worker to determine when that worker receives access to the requester's future tasks. This change gives requesters an incentive to rate workers accurately, because rating a worker a \checkmark+ gives the worker early access to that requester's future tasks and rating a worker a \checkmark- gives the worker late access to that requester's future tasks (Figure~\ref{fig:cascade}). In other words, inflating a ranking so that a poor worker gets a \checkmark\ or \checkmark+ will greatly increase the odds that this worker will ``boomerang'' and return to do more of that requester's tasks, indirectly penalizing the requester.

After a requester posts a new project, the workers who they had previously rated a \checkmark+ will immediately gain access. If the project remains on the marketplace and only a few \checkmark+ workers are accepting it, then the task becomes available to workers with a \checkmark\ rating from that requester, and eventually to workers with a \checkmark-. If a requester has not explicitly rated a given worker, then Boomerang uses that worker's global average rating from all requesters. Within these \checkmark+, \checkmark, \checkmark- groups, we create finer granularity in the cascade by using workers' global average rating as a secondary sort as we release tasks.

This cascaded release of tasks only triggers the next group of lower-rated workers if the current group of higher-rated workers is not accepting the task and completing it rapidly (Figure~\ref{fig:cascade}). To implement this cascaded release, Boomerang measures the current task completion rate using a sliding window and compares the current rate to the maximum completion rate observed with the current minimum rating level. When the current rate is a small enough percentage of the maximum rate, utilization is low and Boomerang reduces the minimum rating and resets its maximum. Intuitively: task completion rates spike immediately following a release to a set of new workers, then tapers off. When the rate is low relative to its initial spike, indicating that workers have tried and abandoned the task, Boomerang releases to additional workers. If the rate remains high after the initial spike, this indicates high utilization---that workers are continuing to engage with the task.

Internally, Boomerang represents a \checkmark- as a score of 1, \checkmark\ as 2, and \checkmark+ as 3. All projects have a minimum rating threshold, representing which worker group can access the projects' tasks. New projects are initialized with a minimum score of 3. The minimum rating threshold is measured with a sliding window period of $T$. Where $t$ is the current time, $t_{init}$ is the time that Boomerang last reduced the minimum score threshold, and $CompletedTasks(start, end)$ is the number of completed tasks between $start$ and $end$, the system measures utilization as the ratio of tasks completed:
\begin{displaymath}
utilization = \frac{CompletedTasks(t-T,  t)}{\max\limits_{t_i \in \lbrack t_{init},\ t-T \rbrack} CompletedTasks(t_i,  t_i+T)}
\end{displaymath}
If $utilization \leq \lambda$ for a utilization ratio $\lambda$ (e.g., 0.3), the Boomerang threshold is reduced to the maximum average rating smaller than the previous threshold value
among the workers who have worked for this requester.

All new Daemo users are seeded with a score just below several \checkmark\ ratings (1.99, where a \checkmark\ scores a 2), which anchors their score with a near-2 ``prior'' so that initial negative ratings do not lock them out of the platform or discourage them from accepting more tasks. To allow workers and requesters to recover from mistakes and inaccurate ratings, we use exponential weighting to calculate each person's rating. Exponential weighting ensures that ratings reflect workers' recent performances and account for quality improvements or declines.
 
\subsubsection{Workers' incentive: task feed ranking}
We considered many possible worker incentives, for example amplifying payment rates for tasks that a worker rates highly, or betting earnings on whether others would also rate a requester highly. However, in our continuous participatory design efforts with workers, it was clear that finding new tasks quickly was a major need that was not well-served by the existing requester-notification tools. Workers spend much of their time searching through thousands of tasks in the task feed and on online forums to identify high-quality, fair-paying work~\cite{chilton2010task, martin2014being, irani2013turkopticon}. Boomerang capitalizes on this need to find tasks by ranking the task feed so that tasks from requesters that a worker gave a high rating are at the top, and requesters receiving low rating are buried at the bottom. Rating good requesters \checkmark+ and bad requesters \checkmark- makes it easier for workers to find tasks they want to do in the future.

As a worker rates requesters, Boomerang determines the ranking of their task feed (Figure~\ref{fig:taskFeed}) as follows. We first form groups of requesters: the requesters that this worker rated \checkmark+ (internal score: 3), those rated \checkmark\ (2), those rated \checkmark- (1), and those brand new to the platform (initialized just below 2). The task feed is then ordered by rating group, with ties broken with each requester's average rating from other workers on the platform. A requester whom a worker rated \checkmark+ and all other workers also rated a \checkmark+ will appear above a requester whom a worker rated a \checkmark+ and other workers rated a \checkmark\ on average. For requesters whom the worker has not yet rated, we use that requester's average rating from other workers as a proxy.
\\
\subsection{Rejection transparency}

Our fieldwork with workers and requesters, as well as prior work~\cite{martin2014being}, indicated three common task rejection patterns for requesters: (1) those who accept all work that they receive; (2) those who reject fairly, i.e.\ when the submission is clearly inappropriate; and (3) those who reject unfairly, i.e.\ even when the work demonstrated clear effort and skill. Requesters in group (1) are easily exploited and lead to an equilibrium in which workers submit low-quality work without fear of rejection. Requesters in (3) take advantage of workers and disenfranchise honest, hard-working individuals.

Requesters hold the power to deny payment without worker contestation on platforms such as Amazon Mechanical Turk. As a result, a worker's decision to select a task is substantially impacted by the likelihood of rejection. For example, Martin \textit{et al}.\ report a worker post on TurkerNation~\cite{martin2014being}: \textit{``Got a mass rejection from some hits I did for them! Talked to other turkers that I know in real life and the same thing happened to them. There rejection comments are also really demeaning. Definitely avoid!''}

\begin{figure}[tb]
\centering
  \includegraphics[width=1.0\columnwidth]{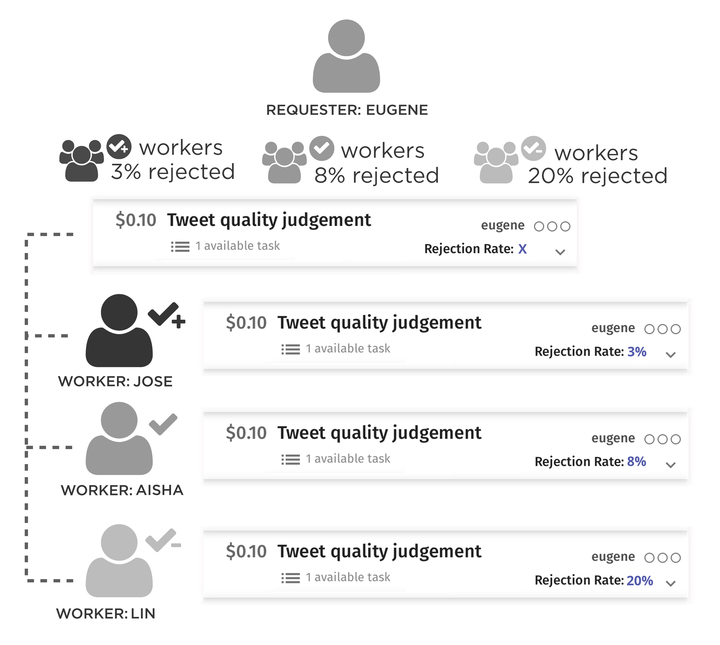}
  \caption{Workers see a personalized rejection rate for every requester on their task feed. By leniently or frequently rejecting work of various qualities, a requester influences whether low- or high-quality workers will complete their tasks. Requesters hope to avoid strategic workers from accepting their tasks if they are too lenient or avoiding their tasks if they are too harsh.}
  \label{fig:rejectionmodel}
\end{figure}

Boomerang helps address this issue by displaying each requester's rejection rates on the task feed for tasks submitted by workers like the current user (Figure~\ref{fig:rejectionmodel}). Workers can then use this estimate of their own rejection rate as an aid in the task selection process. In other words, a low-rated worker would see the requester's rejection rate for tasks from low-rated workers, and a high-rated worker would see the requester's rejection rate for tasks from high-rated workers. Boomerang does not display a single rejection rate for the requester, because it is too distanced from each worker's experience to sway the worker's behavior, and would allow a requester to unfairly reject a few high-quality workers and accept many low-quality workers just to maintain an acceptable rate and not accept and reject for legitimate reasons.

Requesters who are aware that their rejection rate will be visible have an incentive to rate accurately. If they blindly accept all work from workers in a given rating category (\mbox{\checkmark-,} \mbox{\checkmark,} or \mbox{\checkmark+}), they may attract workers who are strategically seeking tasks that are unlikely to get rejected even if they are less accurate than usual. However, if the requester begins rejecting too much work, workers will avoid taking the task unless they are willing to risk getting a lemon~\cite{akerlof1970market}.

To calculate a requester's rejection rate for a worker to view, we partition the range of reputation scores (1--3) into six equal buckets and select the bucket that contains the worker's rating. We calculate a weighted average of the percentage of that worker's submissions that the requester has rejected and the percentage of work that the requester has rejected submitted by everyone else in the worker's bucket. We favor the individual worker while computing this weighted rejection rate. If the worker has not completed any work for the requester, we use only the rejection rate of workers in their bucket as a proxy for their rejection rate. And if the requester has not rejected any tasks yet, we use the rejection rate of the worker's bucket by requesters with similar reputation. 

\section{Evaluation}
This paper draws on incentive compatibility to inspire designs for more honest reporting on crowdsourcing platforms. Traditionally, evaluating incentive-compatible mechanisms requires an economic game with rules that can be modeled via a payoff function, and requires a mathematical proof that either 1) deviating from honest reporting is never better and sometimes strictly worse than telling the truth no matter what others do (\emph{strategyproofness}), or 2) if everyone else reports honestly, it is optimal for the agent to also report honestly (a \emph{Nash equilibrium})~\cite{nisan2007algorithmic}. However, our goal is to engage not in formal mechanism design, but in the wicked problem of socio-technical interaction design, which operates at a level of complexity and scale where it is impossible to fully model everyone's behavior or payoffs~\cite{rittel1973dilemmas}. With this wicked problem, the outcome of these interventions depends on how strongly each incentive will motivate each user. So, in our evaluations, we opt for a behavioral outcome via a field experiment instead of a theoretical guarantee. This allows us to measure the empirical strength of each design intervention.

We performed two studies: the first tested the Boomerang reputation feedback design, and the second tested the Boomerang task rejection design. Each of these studies was performed twice, once for requesters and once for workers. 

\subsection{Study 1: Reputation feedback}
Our first study examined whether the Boomerang reputation system produced more honest feedback from requesters and workers.

\subsubsection{Requesters: Method}
We conducted a between-subjects field experiment to test the effectiveness of Boomerang in incentivizing requesters to provide more accurate ratings of workers. To do so, we compared requesters' rating feedback to a forced-choice evaluation of each worker's quality.

We recruited 35 requesters who had experience on a microtask crowdsourcing platform and randomized them between the Boomerang and control conditions. The population included professionals and academic requesters, with a skew toward academics. The requesters performed a first phase of rating, then a second phase of forced-choice decisions.

For their rating phase, requesters were shown a series of three representative yet difficult microtasks in sequence: video summarization, fact extraction from a webpage, and webpage viewpoint classification. Prior to the study, we pre-populated the results by recruiting 21 workers from Amazon Mechanical Turk to complete every task. We recruited these workers with a variety of different HIT completion and acceptance requirements to ensure a distribution of worker quality. However, requesters did not see all 21 workers' results for every task: for each task, we sampled seven workers and showed only their results to the requester, reserving the other workers as ``the crowd'' that Boomerang might allow or prevent from taking future tasks based on the feedback. Requesters saw results from each task one at a time and were asked to rate workers based on the quality of their results.

Requesters in the control condition saw a traditional rating interface: ``I like this'', ``Default'', and ``I don't like this''. These responses did not impact which workers performed upcoming tasks. In the Boomerang condition, the interface stated: ``Like: Grant this worker early access to my tasks, so they do a lot of my work'', ``Same: Give this worker access at the same time as normal workers, so they do a bit of my work'', and ``Bury: prevent this worker from getting access to my tasks until last, so they rarely do my work''. In both conditions, the seven workers (out of 21) who performed the first task were sampled uniformly from the full set. In the control condition, the workers for the subsequent two tasks were likewise uniformly sampled from the full set. In the Boomerang condition, we simulated the effects of the cascading task release by weighting workers' probability of being chosen for future tasks. Workers with a high rating were given double the baseline probability of being sampled, workers with a middle rating were left at the default rate, and workers with a low rating were given half the baseline probability of being sampled.

After completing the worker ratings, requesters concluded by making a series of ten forced-choice decisions between sets of three workers each. In each set, requesters answered which of the three workers they felt was highest-quality. These forced-choice decisions captured requesters' private opinions at a more fine-grained level, enabling us to compare their private decisions to the ratings that they provided earlier. We randomly sampled sets of three workers who had completed at least one task for the requester. When making this final decision requesters could see all of the results that each worker had been sampled to complete. It was important that requesters not feel that their decisions here would impact their later experience, to ensure comparable decisions between conditions. Thus, no Boomerang explanations appeared on these ratings. In contrast, whenever a decision in the first phase carried a Boomerang impact, this was clearly stated in the user interface. 

We measured how closely requesters' forced-choice decisions matched their ratings. To do so, we used the \checkmark-, \checkmark, and \checkmark+ ratings to predict requesters' ten forced-choice decisions. We calculated a \emph{score}, or the expected value of the number of correct predictions. For example, if a requester had ranked two workers as \checkmark\ and one as \checkmark+, we predicted that they would choose the \checkmark+ worker. If correct, the number of expected correct answers rose by one. If there were ties, e.g.\ two workers were rated \checkmark, the number of expected correct answers rose by $\frac{1}{2}$ and $\frac{1}{3}$ if a three-way tie. In other words, the score was the expected value of the number of correct answers on the forced-choice task using the ratings as the only source of information. During analysis, we dropped participants who indicated low attention by having three or more inconsistent decisions in the forced-choice task (i.e.\ preferring one worker over another in one round, then swapping the opinion in another round), or as is common, were statistical outliers of at least three standard deviations from the mean. We then performed an unpaired t-test to compare the scores between the two conditions. We hypothesized that requesters in the Boomerang condition would have higher scores --- in other words, ratings that were more predictive of their forced-choice decisions.

\begin{figure}[tb]
\centering
  \includegraphics[width=1.0\columnwidth]{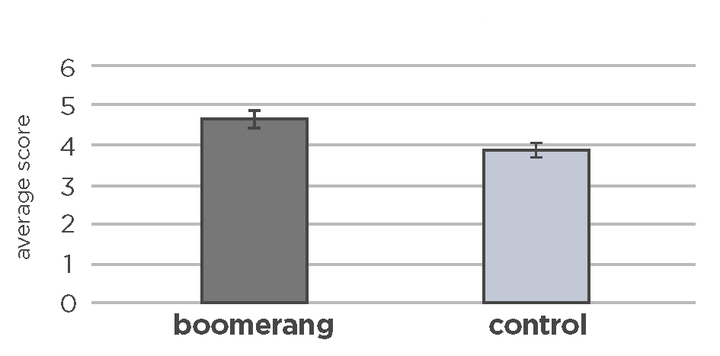}
  \caption{Requesters' average scores in the Boomerang condition was higher than the control condition, indicating that Boomerang was more effective at producing informative ratings: t(33)=2.66, p<.01, Cohen's d= 0.91. }~\label{fig:reqavgscorerating}
\end{figure} 

\subsubsection{Requesters: Results}
Requesters in the control condition had an average score of 3.9 out of 10 ($\sigma=.79$), and requesters in the Boomerang condition had an average score of 4.6 out of 10 ($\sigma=.83$). An unpaired t-test confirmed that requesters in the Boomerang condition scored significantly higher: $t(33)=2.66$, $p<.01$, Cohen's $d=0.91$. Thus, Boomerang had a large positive effect on  the information contained in ratings . Moreover, the rating distribution was sharply shifted: requesters in the control condition reported 15.2 \checkmark+ ratings out of 21 on average, but in the Boomerang condition, this dropped by nearly half to 8.7 \checkmark+ ratings on average. The Boomerang condition's de-inflated ratings shifted into \checkmark\ (7.6, vs. 3.1 in control), and into \checkmark- (4.6, vs. 2.6 in control). Boomerang markedly redistributed requesters' ratings, as the \checkmark+ rating was used sparingly, resulting in more informative reputation feedback.

\begin{figure}[tb]
\centering
  \includegraphics[width=1.0\columnwidth]{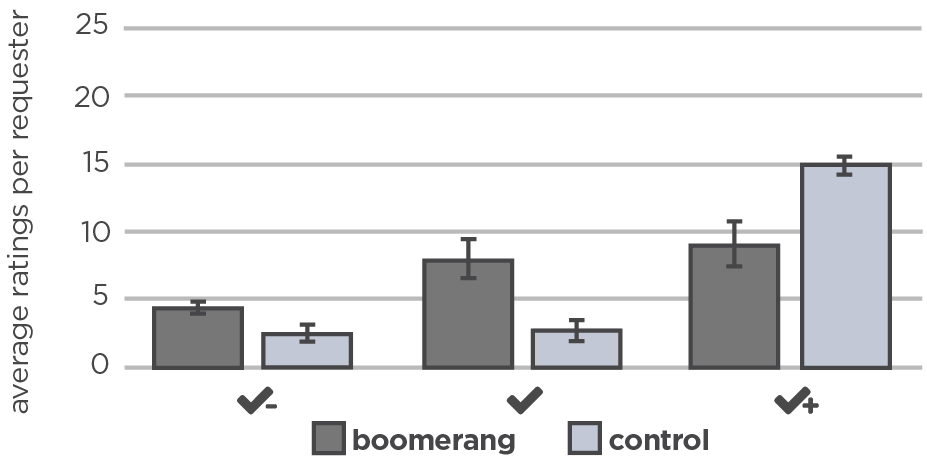}
  \caption{Boomerang markedly redistributed requesters' ratings, as the \checkmark+ rating was used more sparingly.}~\label{fig:perreqscore}
\end{figure}

\subsubsection{Workers: Method}
As with the requesters, we used a between-subjects study to test the effectiveness of Boomerang's feedback system in incentivizing workers to provide more accurate ratings to requesters. We similarly predicted that Boomerang would yield ratings closer to workers' forced-choice decisions about requesters than a control rating interface.  

The setup of this study mirrored that of the requester study. We recruited 42 workers from Amazon Mechanical Turk who had over 500 HITs completed and over a 95\% acceptance rate. The study lasted roughly an hour, so we paid workers \$10 in line with current ethical standards on Mechanical Turk~\cite{salehi2015we}. We randomized the workers between Boomerang and control conditions.

To create tasks for the workers to perform, we sampled over fifty tasks from Mechanical Turk and recreated them exactly on Daemo. We sampled these tasks to canvas a variety of high-, medium-, and low-rated requesters based on their Turkopticon ratings~\cite{irani2013turkopticon}. We then evenly distributed these fifty tasks among seven requesters, so that each requester would have tasks of matched quality.

Workers were told that they would be completing tasks in two phases: the first would consist of a small number of tasks with no time limit, and the second would consist of a 15-minute time limit to complete the best tasks in a very large set of tasks. Because workers did not have a long-term stake in their experience on Daemo, we could not rely on long-term feed curation as a motivator by itself: we thus used the looming larger set of tasks to motivate workers to care about quickly uncovering high-quality tasks. Workers then entered the first phase to complete fourteen tasks, two per requester. They rated each task as they proceeded.

In the control condition, the rating feedback was similar to traditional reputation systems, with ``I like this'', ``Default'', and ``I don't like this''. The feedback did not impact future task feed ranking. In the treatment condition, the feedback communicated the Boomerang mechanism: ``I like this: feature this requester's tasks at the top of my task feed in the future'', ``Same: keep this requester's tasks in the middle of my task feed'' and ``I don't like this: bury this requester's tasks at the bottom of my task feed in the future''. Workers in the treatment condition saw their task feed re-rank as they entered this rating feedback.

After workers finished all fourteen tasks and provided ratings for them, we introduced a series of private forced-choice decisions between requesters. These forced-choice decisions captured workers' opinions at a more fine-grained level, enabling us to compare their private decisions to the ratings that they provided previously. Each forced-choice decision displayed three randomly selected requesters from the set of seven, as well as the tasks that the worker had completed for each of the three requesters. Workers were asked to select which of the three requesters they preferred. Workers' previous rating feedback was hidden during this phase. As before, no Boomerang-related information appeared in the forced-choice comparisons, which were designed to communicate that workers' opinions were private and had no impact on later phases. We repeated this forced-choice ten times by re-sampling sets of three requesters. Finally, when workers arrived at the second phase of the task, we told them that they had been selected to skip the second phase of the task and would be paid in full. Workers believing that the second phase of the task was coming was sufficient to set their motivation; completing the second round of tasks was unnecessary once they actually arrived.

\begin{figure}[tb]
\centering
  \includegraphics[width=1.0\columnwidth]{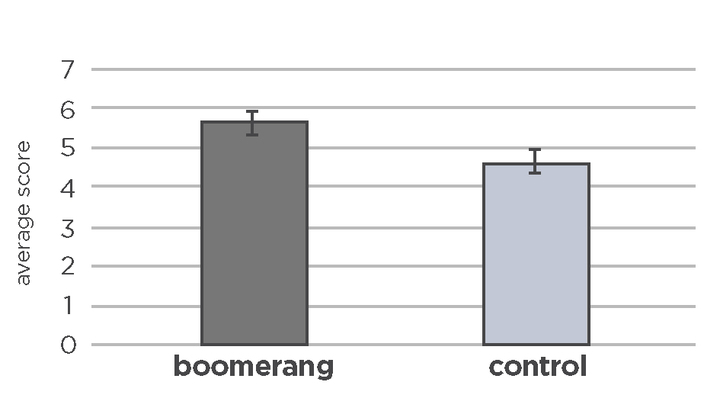}
  \caption{Workers' average scores in Boomerang was higher than the scores in the control condition, indicating that Boomerang was more effective at producing informative ratings: t(40)=2.87, p<.01, Cohen's d= 0.89.  }~\label{fig:workeravgscore}
\end{figure}

Similar to the requester study, we measured how closely workers' forced-choice decisions matched their ratings. To do so,  we calculated an expected number of correct predictions given the ratings from the first phase to predict the decisions in the second phase. We filtered out outliers and inattentive participants using the same techniques as the requester study, then used an unpaired t-test to compare the scores between workers in the control and Boomerang conditions. We hypothesized that workers in the Boomerang condition would have higher scores --- in other words, that Boomerang ratings would be more predictive of workers' forced-choice decisions.

\subsubsection{Workers: results}
Workers in the control condition had an average score of 4.5 out of 10 ($\sigma=1.3$), and workers in the Boomerang condition had an average score of 5.6 out of 10 ($\sigma=1.6$). An unpaired t-test confirmed that workers in the Boomerang condition scored significantly higher: $t(40)=2.87$, $p<.01$,  Cohen's $d=0.89$. Mirroring the result for requesters, Boomerang produced a large positive effect on score. Again this outcome was a result of a shifted rating distribution: workers in the control condition reported 3.5 \checkmark+ ratings out of seven on average, but in the Boomerang condition, this dropped by nearly 15\% to 3.0 \checkmark+ ratings on average. To compensate, Boomerang condition workers increased their number of \checkmark- ratings by 11\% (average 1.8 vs. 2.0) and their number of \checkmark ratings by 19\% (average 1.7 vs. 2.0). So, Boomerang had an effect of de-inflating the ratings that workers gave, as the \checkmark+ rating was used more sparingly.

\begin{figure}[tb]
\centering
  \includegraphics[width=1.0\columnwidth]{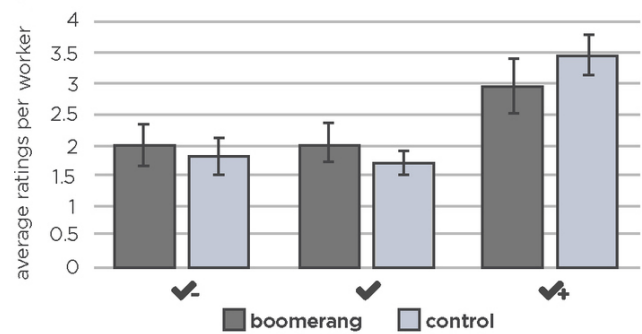}
  \caption{Boomerang had an effect of de-inflating the ratings that workers gave, as the \checkmark+ rating was used more sparingly.}
  \label{fig:wrkravgscorerating}
\end{figure}

\subsubsection{Measurement validation}
During the study, workers and requesters made private forced-choice decisions that captured their detailed opinions, which we used to compare the Boomerang and control ratings. In this setup, it is important to verify that participants in the Boomerang condition did not feel that their forced-choice decisions would impact them later, or else the two conditions might not be comparable. To check this assumption, we further examined our data. 

First, if Boomerang participants felt that their private forced-choice decisions would impact their later experience, then forced-choice decisions made in the control vs Boomerang should differ even when faced with the exact same comparison sets. We compared participants' decisions when faced with the same forced-choice decision (e.g., choosing between $A$, $B$, $C$) in the control and Boomerang conditions. To reduce data sparsity, we translated each 3-way choice into paired comparisons (choosing $A$ over $B$ and $C$ implies $A>B$ and $A>C$). For each pair ($A$ and $B$, $A$ and $C$, $B$ and $C$, etc.), we calculated the percentage of participants who chose the more popular of each pair, e.g., 81\% Boomerang vs 71\% control. If rating behavior differed between conditions, we would expect Boomerang decisions to more reliably identify the better requesters (thus, have higher percentages), and control condition decisions to be more random (lower percentages) because they have less incentive to attend to the decision. We performed a paired t-test between the two conditions with each pair forming an observation. The t-test is not significant (t(14)=1.7, n.s.), suggesting that the conditions did not rate differently on the same decisions, and the two conditions are thus comparable.

We complemented this analysis with a second investigation: if the two conditions produced different rating behavior in the ground truth forced-choice evaluations, their decisions would agree at different rates  when benchmarked against a neutral rater's decisions for the same forced-choice options. We thus performed the forced-choice decisions ourselves, blind to the original decisions and condition. Two raters independently coded answers for each forced-choice decision, with a third rater resolving disagreements. Because these choices are subjective, neither condition agrees completely with the neutral ratings. However, they agree at roughly the same rate with the neutral ratings (62\% vs 65\%), again suggesting no bias in forced-choice behavior between conditions, and that these decisions likely captured honest private opinions.

\subsection{Study 2: Rejection rates}
The first study demonstrated that Boomerang produces feedback that is more tightly aligned with workers' and requesters' opinions. In this study, we generalize the result by performing a between-subjects experiment to measure the behavior shift in response to Boomerang's rejection visibility. 

\subsubsection{Requesters: Method}
We began by obtaining results for requesters to review. We used Amazon Mechanical Turk workers' responses to 24 title generation tasks~\cite{bernstein2012direct} that asked workers to transform a series of queries into a descriptive title that summarizes the queries. Workers' responses were hand-labeled as correct or incorrect by two raters, and a third rater resolved any differences. We use these hand-labeled decisions as ground truth, and measured whether requesters in the Boomerang condition were more likely to produce the same decisions.

We recruited 12 requesters who had experience posting tasks on Amazon Mechanical Turk and randomized them between Boomerang and control groups. We provided each requester with the instructions that workers received, as well as the results, and asked them to review the submissions to accept or reject as they would on Mechanical Turk. Requesters in the control condition made their decisions without additional information, as they would on traditional platforms. Requesters in the Boomerang condition were shown instructions that explained to them how Boomerang's rejection visibility worked, including how low- and high-quality workers would see the rejection information in their task feeds. Workers' reputation scores were not shown to requesters during this experiment.

We measured each requester's accuracy as the number of accept or reject decisions they made that matched the ground-truth decisions. We performed an unpaired t-test to compare the accuracy of requesters in each condition. We hypothesized that requesters in the Boomerang condition would have higher accuracy than those in the control condition. 

\subsubsection{Requesters: Results}
The hypothesis was supported: requesters in the Boomerang condition were more accurate in matching the pre-labeled correct/incorrect decisions in their acceptance and rejection decisions. The mean accuracy in the Boomerang condition ($78\%$, $\sigma=6\%$) was higher than the mean accuracy in the control condition ($66\%$, $\sigma=11\%$). An unpaired t-test comparing the two conditions confirmed that the difference was significant: $t(10)=2.235$, $p<0.05$, Cohen's $d=1.29$

\subsubsection{Workers: Method}
Finally, we sought to measure whether displaying rejection rates in the task feed would cause workers to actually seek out or avoid tasks, as requesters were led to believe. We thus recruited 40 workers from Amazon Mechanical Turk. Each worker was paid a base rate of \$10. We asked workers to complete tasks for fifteen minutes and make as much money as possible on the platform. We offered to pay them a bonus equal to their earnings on Daemo, provided that the requesters for each task accepted their work. For this time period, the task feed was populated with over fifty tasks used in Study 1, grouped across fourteen requesters. We randomized each requester's rejection rate into three buckets: low (0-3\%), medium (10-20\%), and high (40-60\%). These rejection buckets were empirically determined with feedback from Mechanical Turk workers on forums. The result was a series of tasks of varying quality (per the original Turkopticon ratings), augmented with a randomized set of rejection rates. 

We then measured the number of tasks completed in each category (low, medium, high) by the workers. We performed a chi-square test to compare the number of tasks completed in each category.

\subsubsection{Workers: Results}
Workers were inclined to select tasks from requesters with low rejection rates (i.e., 0\%-3\%). Out of all 180 tasks completed, 75\% of the tasks done were of low rejection rate requesters. The rest were from medium (11\%) and high (14\%) rejection rate requesters. A chi-square test confirmed that this difference was statistically significant, $\chi^2(2)=141.0$, $p<.001$.

\subsection{Summary}   
Study 1 showed that requesters' and workers' average ratings in the Boomerang condition closely matched their private opinions, resulting in more informative and less inflated reputation ratings. Study 2 demonstrated that workers react to the visibility of rejection information by selecting the tasks from requesters with a low rejection rate, and requesters become more accurate at accepting and rejecting submissions when they know this is the case.

\section{Discussion}
In this section, we reflect on the methodological limitations of our studies, as well as the broader design space of incentive-compatible interaction design. Our studies suggest that setting incentive compatibility as a design goal can translate to more accurate information on socio-technical platforms.

\subsection{Limitations}
The incentives in our design assume that users remain in the social system for long periods of time and thus have a stake in making the social system useful. An ideal evaluation would involve workers and requesters who hold such a stake, either by manipulating Amazon Mechanical Turk (which remains inaccessible), or recruiting long-term workers and requesters onto the Daemo platform. Because Daemo is still nascent, however, we opted for a field experiment on Daemo that used workers and requesters from Mechanical Turk. This decision does limit the external validity of the experiments. However, because it minimized participants' long-term motivations, it is also biased toward a conservative estimate of the strength of the effect. There are also possible novelty effects, which will need to be teased out in future work via longitudinal studies. Finally, while we tried to recruit a wide variety of crowd workers and requesters and used real tasks from crowd marketplaces, our results might not yet generalize to crowd work at large.

Not every user will be incentivized to rate carefully with Boomerang --- it is not a forcing function, but a (strong) nudge for the rational actor. One challenge is that requesters may not have the time or inclination to rate potentially hundreds of workers for each task. In the deployed version of Daemo, we thus only require requesters to rate a small, randomly sampled set of workers (5--10) for every task. The motivation is that this stochastic rating, especially if it biases toward workers with less coverage, generates good rating coverage for workers: assuming a worker performs twenty different microtasks (e.g., image labeling) a day, and requesters are on average rating even 5\% of their worker pool, each worker will still receive at least five ratings in a typical work week. A second challenge is that a requester who plans to have only one task completed on the platform may see little value in rating when the outcomes relate only to hypothetical future workers. In the inverse situation, since the longevity of requesters is often variable, workers may not see any incentive in providing ratings if the requester will potentially never post work again. We aim to clearly communicate how honest ratings will improve the overall quality of task completion on the platform.

\subsection{The design space of incentive-compatible interaction}
\begin{figure}[tb]
\centering
 \includegraphics[width=1.0\columnwidth]{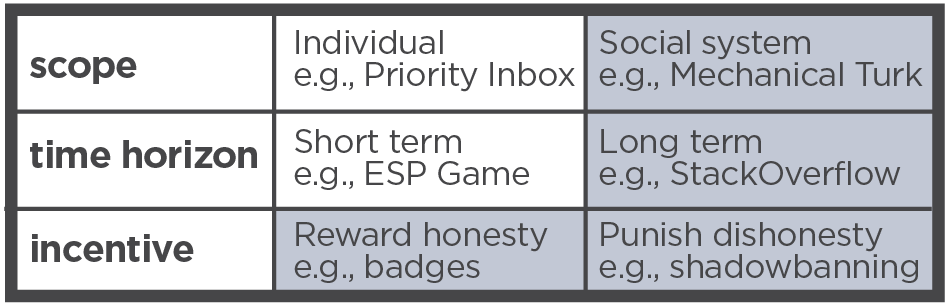}
 \caption{The design space of incentive-compatible interaction design. Blue cells represent areas explored by Boomerang.}
~\label{fig:F-Discussion}
\end{figure}

What is the broader design space of incentive-compatible interaction design and how do we operationalize it? Figure~\ref{fig:F-Discussion} represents our reflection to inform future designs.

First, \textit{scope} indicates the number of people engaged: either an individual (and the system), or an entire social system. With an individual, for example with Gmail's Priority Inbox, there is no negotiation with other people --- rating an email as important provides better filtering for you in the future. With a social system, the interactions are directly engaging with other individuals, as in crowdsourcing workers interacting with requesters. We suggest that incentive-compatible interaction design becomes more important in social environments, because participants have multiple competing goals. Second, \textit{time horizon} specifies how quickly the user's decision impacts them. Boomerang is relatively long-term, since the impact does not occur until the next time a requester posts a task. The ESP Game, on the other hand, is immediate---guessing incorrectly lowers both the user's potential point total and their partner's. We hypothesize that the shorter term the impact, the more effective the design. Finally, \textit{incentive} specifies that interventions can either reward honesty, as in badges or additional features, or punish dishonesty, as in connecting dishonest users only with other dishonest users (shadowbanning). Boomerang's reputation system seeks to do both---honesty makes good work more likely, whereas dishonesty makes bad work more likely---but the rejection design mostly only punishes dishonesty. As long as truth-telling behavior is a dominant strategy (i.e.\ it produces better outcomes than any other strategy), both approaches can work.

\subsubsection{Implications for other socio-technical platforms}
How might other social computing platforms apply incentive-compatible interaction design? Much like Boomerang targets reputation systems in a marketplace, other social platforms could target feedback and reputation systems as well. For example, Airbnb hosts' ratings for guests could influence which users later see the their home at the top of Airbnb's search results: a host who gives a guest with kids a high rating may appear higher in the search rankings for future users traveling with kids. A second, more general category includes incentivizing pro-social behavior. For example, Uber could give lower priority to riders who cancel often. Likewise, Waze could reward users who give reports on app by granting them early access to others' new information. 
 
With respect to crowdsourcing platforms in particular, we have focused thus far on reputation systems (e.g. ratings, rejection). However, incentive-compatible interaction design could be applied to many other aspects of the platform. For example, one of the main concerns workers have is whether a given task will pay enough in practice to meet their minimum fair wage~\cite{martin2014being}. A crucial piece of information required to identify such tasks is an accurate estimate of the time required by the worker to complete it. This information is not readily measurable because workers have no incentive to report this number after they complete a task, automatic logs are not accurate~\cite{jeffrz2011UIST,ChrisCallison2014crowdworkers,Justin2015ETA}, and requesters underestimate how long a task will take~\cite{hinds1999curse}. 

\begin{figure}[tb]
 \includegraphics[width=1.0\columnwidth]{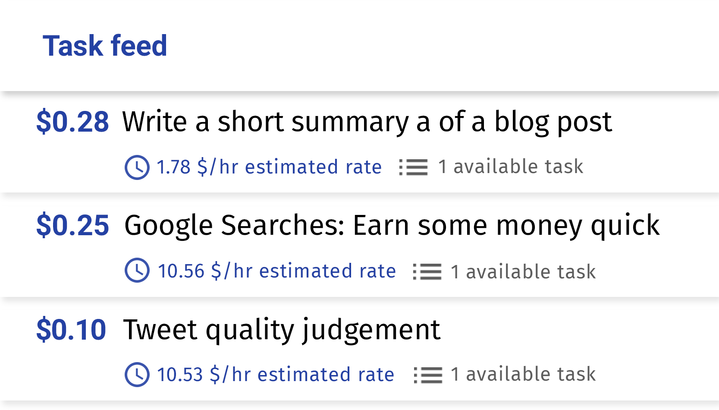}
 \caption{By asking workers to measure and accurately self-report how long tasks take them, Boomerang can estimate workers' hourly wage for other tasks on the platform.}
~\label{fig:designspace}
\end{figure}

An incentive-compatible interaction design might incentivize a worker to share the accurate time. To prototype this, we have added an optional timer to each task in Daemo, then ask workers to edit its recorded time. In exchange, Daemo uses the information to predict that worker's hourly wage for all other tasks on the platform (Figure~\ref{fig:designspace}). To do so, Daemo estimates a multiplier of how much faster or slower than the average each worker completes tasks. With this information, any task on the platform with at least one worker self-reporting their time can show an estimated time to all other  workers. (All new tasks on Daemo go through an initial ``prototype'' phase where a small number of workers self-report completion time~\cite{gaikwad2015daemo}, so in practice all tasks on the platform will appear with estimated times.) Workers who report inaccurate times will see inaccurate hourly wage estimations on their task feeds. Therefore, if a worker consistently underestimates or overestimates time spent on tasks, they will receive a task feed that displays skewed estimates, rendering their task selection process more problematic. 

\section{Conclusion}
In this paper, we demonstrate that incentive-compatible interaction design encourages users of crowdsourcing platforms to report private information more accurately. We introduce Boomerang, an incentive-compatible interaction design for a reputation system that aligns the incentives of workers and requesters in reputation ratings and task rejection. Boomerang's task feed uses ratings from both parties to determine which workers will receive early access to the requester's future tasks, and which requester's tasks will appear at the top of the worker's task feed. In our studies, we found that Boomerang's design led to more accurate reputation ratings for workers and requesters, and to a task rejection pattern that more accurately reflects the requesters' opinions of workers' submissions.                 

The features introduced by Boomerang and the behaviors that these features are able to motivate both workers and requesters, demonstrating that incentive-compatible interaction design may help prevent crowdsourcing platforms from becoming markets for lemons.

\section{Acknowledgements}
We thank the following members of the Stanford Crowd Research Collective for their contributions: Shivam Agarwal, Shrey Gupta, Kevin Le, Mahesh Murag, Manoj Pandey, Akshansh, Shivangi Bajpai, Divya Nekkanti, Lucas Qiu, Yen Huang, Vrinda Bhatia, Yash Sherry, Paul Abhratanu, Varun Hasija, Aditi Mithal, Justin Le, Sanjiv Lobo, Xi Chen, Karthik Chandu, Jorg Doku, Prithvi Raj, Archana Dhankar, Anmol Agarwal, Carl Chen, Armin Rezaiean-Asel, Gagana B, Jaspreet Singh Riar, Raymond Su, Vineet Sethia, Klerisson Paixao, Ishan Yeluwar, Yash Mittal, Haritha Thilakarathne, Alan James Kay, Zain Rehmani, Anuradha Sharma, Aalok Thakkar, Prabath Rupasinghe, Kushagro Bhattacharjee, Nikita Dubov, Yashovardhan Sharma, and Namit Juneja. We also thank requesters, workers, and Chris Callison-Burch for their help with the study. This work was supported by NSF award IIS-1351131 and a Stanford Cyber-Social Systems grant.

\balance

\bibliographystyle{acm-sigchi} 
\bibliography{uist_paper}
\end{document}